\documentclass{elsart3p}

 \usepackage{graphicx}

\usepackage{amsmath,amssymb}

\begin{document}

\begin{frontmatter}



\title{Thermal Effects in the dynamics of disordered elastic systems}


\author[bar]{S. Bustingorry}
\author[bar]{A. B. Kolton}
\author[ors]{A. Rosso}
\author[ens]{W. Krauth}
\author[gen]{T. Giamarchi}
\address[bar]{Centro At\'omico Bariloche, 8400 S.C. de Bariloche, Argentina}
\address[ors]{CNRS; Univ. Paris-Sud, UMR 8626, ORSAY CEDEX, F-91405, LPTMS, France}
\address[ens]{CNRS-Laboratoire de Physique Statistique, Ecole Normale
Sup{\'{e}}rieure, 24 rue Lhomond, 75231 Paris Cedex 05, France}
\address[gen]{DPMC-MaNEP, University of Geneva, 24 Quai Ernest Ansermet, 1211 Geneva 4, Switzerland}

\begin{abstract}
Many seemingly different macroscopic systems (magnets, ferroelectrics, CDW, vortices,..) can be described as generic disordered elastic systems.
Understanding their static and dynamics thus poses challenging problems both from the point of view of fundamental physics and of practical applications.
Despite important progress many questions remain open. In particular the temperature has drastic effects on the way these systems respond to an external force. We address here the important
question of the thermal effect close to depinning, and whether these effects can be understood in the analogy with standard critical phenomena, analogy
so useful to understand the zero temperature case. We show that close to the depinning force temperature leads to a rounding of the depinning transition
and compute the corresponding exponent. In addition, using a novel algorithm it is possible to study precisely the behavior close to depinning, and to show that the commonly accepted analogy
of the depinning with a critical phenomenon does not fully hold, since no divergent lengthscale exists in the steady state properties of the line below the depinning threshold.
\end{abstract}

\end{frontmatter}

\section{Introduction}
\label{sec:intro}

Disordered elastic systems (DES) are ubiquitous in condensed matter. They are in particular relevant for
magnetic~\cite{lemerle_domainwall_creep,bauer_deroughening_magnetic2,yamanouchi_creep_ferromagnetic_semiconductor2,metaxas_depinning_thermal_rounding} or ferroelectric~\cite{paruch_ferro_roughness_dipolar,paruch_ferro_quench} domain walls, contact lines~\cite{moulinet_distribution_width_contact_line2}, fractures~\cite{bouchaud_crack_propagation2,alava_review_cracks},
vortex lattices~\cite{blatter_vortex_review,du_aging_bragg_glass}, charge density waves~\cite{nattermann_cdw_review} or Wigner crystals~\cite{giamarchi_electronic_crystals_review}.
In these systems, one particularly important question is to understand the response of the interface
to an externally applied force, such response being directly measurable in all the above systems \cite{giamarchi_domainwall_review}.
This problem is an extremely complicated one, since the competition between elasticity and pinning leads to the presence of many metastable states and thus to glassy properties.

Despite this inherent complexity, we start now to understand well the dynamical properties of such systems at zero temperature.
In particular, the system remains pinned up to a critical force $F_c$ and starts sliding above this force. A particularly fruitful approach
was to view this depinning transition in the light of critical phenomena~\cite{fisher_depinning_meanfield},
where the external force $F$ would act as a control parameter analogous
to the temperature leading to the transition. The velocity $v$ would be the analogous of the order parameter, finite on one side of the transition
and zero on the other side. This analogy immediately suggests the presence of critical exponents such as the ones defining the order parameter
$v \propto (F-F_c)^\beta$ and the existence of a divergent lengthscale $\xi \propto (F-F_c)^{-\nu}$, that can be in that case identified with the
avalanche size when the system starts to move. These intuitive ideas are well confirmed in numerical and analytical studies of the $T=0$ systems.
Given the success of the intuition gathered from the analogy with the critical phenomena, one is tempted to extend it to the case of finite temperature. In that case, thermal activation always allows to pass barriers ensuring a finite velocity at any force. Close to the depinning force
$F_c$ the finite temperature would thus be naively analogous to an external field (such as a finite magnetic field) coupling to the order parameter
(e.g. the magnetization in a magnetic system) and ensuring that this parameter is finite on both sides of the transition. This analogy suggests
immediately two important consequences: a) there must exist a rounding of the depinning transition due to the temperature and a thermal rounding
exponent $V(F_c)\sim T^\psi$, as for standard critical rounding; b) a divergent lengthscale must exist on \emph{both} sides of the
transition, and at ``short'' lengthscales the system must resemble the ``critical'' system, while above this divergent length the system must
reflect the corresponding fixed point (high or low temperature for a normal critical phenomenon, zero force or high force in the case of the
depinning transition). These consequences for the finite temperature behavior are quite striking and of course of direct experimental relevance,
since experiments are always performed at a finite temperature.

However, despite the success of the analogy with critical phenomena for $T=0$, its extension to finite temperature is not guaranteed.
It is thus important to check what remains of these properties. We address here these two points, and show
that while a thermal rounding does indeed exist, there is no divergent lengthscale associated with the steady state properties of the line
for $F< F_c$, showing that the analogy with a standard critical phenomenon does not hold and making it a challenge to develop an intuitive picture
of the depinning transition at finite temperatures.

\section{Thermal rounding} \label{sec:thermal}

We first concentrate on the question of thermal rounding.

\begin{figure}
\includegraphics[width=7cm]{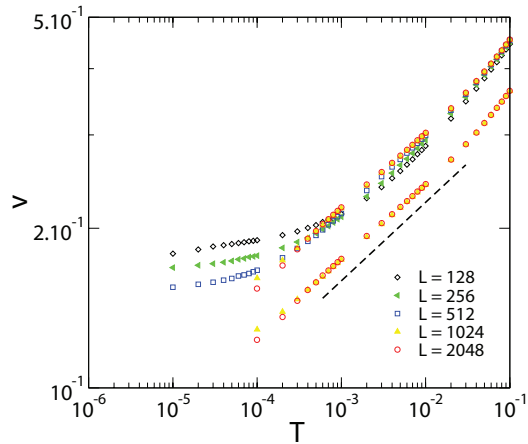}
\caption{\label{fig:rounding} Steady state velocity against temperature for different system sizes. For large enough systems ($L \ge 1024$), all finite size effects are suppressed
in the range of temperature measured. This can be seen by comparing the data for $L=1024, 2048$, that has
been offset for clarity. The dashed line indicates the $v \sim T^\psi$ behavior, with the thermal exponent $\psi \simeq 0.15$ fitted using only $L=1024, 2048$. After \cite{bustingorry_thermal_rounding}.}
\end{figure}

Although analytical equations for the motion at finite temperature do exist \cite{chauve_creep} they are in practice very complicated
to solve, and close to depinning no definite picture has emerged from them yet. However this question can be tackled by a numerical
analysis of the equations of motion. We used a new technique allowing to avoid sample to sample fluctuations of the pinning force, which
allowed us to obtain a very good statistics and to extract the depinning exponent. Details and further reference to previous studies
of this problem in the literature can be found in Ref.~\cite{bustingorry_thermal_rounding}. Fig.~\ref{fig:rounding} shows the steady state velocity of a system
at the critical force.

Data shows a good scaling of the form $v \sim T^{\psi}$, leading to the value $\psi \simeq 0.15$. More importantly
one could show by measuring the structure factor of the interface that both the thermally-rounded and the $T=0$ depinning, display the same large-scale geometry, described by an identical divergence of a characteristic length $\xi$ with the velocity of the form $\xi \propto v^{-\nu/\beta}$, where $\nu$ and $\beta$ are respectively the $T=0$ correlation and depinning exponents.

This remarkable fact fits very well universal scaling ideas from standard critical phenomena. It is important also to point out that so far there is no theoretical explanation of the value of the
thermal rounding exponent (see Ref.~\cite{bustingorry_thermal_rounding} for a detailed discussion on that point). Quite paradoxically the best estimate seems to be
provided by a direct application of the formula for static critical phenomena, which of course has no justification in the case of such dynamical
transition. From a more fundamental point of view the scaling found above strongly suggests that the rounding of singularities that appear
in the renormalization study of the depinning \cite{chauve_creep} are still controlled at finite temperatures by the velocity alone, as was the
case for the $T=0$ depinning.

The analysis of thermal rounding would thus comfort the idea to extend the analogy with standard critical phenomena to finite temperatures.
As we will see this is not the case when the characteristic lengthscales of the line are examined.

\section{Characteristic lengthscales} \label{sec:lengths}

\begin{figure}
\includegraphics[angle=-90,width=8cm]{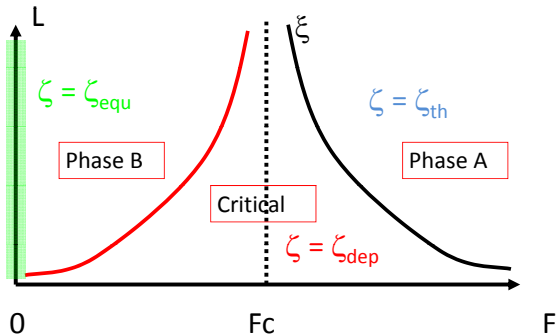}
\caption{\label{fig:regimes} The naive vision of the characteristic lengthscales and roughness of the line, in its moving frame. The analogy with critical phenomena
would suggest a divergent lengthscale on both sides of the transition and a critical regime. This is in very different from what actually happens and is depicted in Fig.~\ref{fig:nodiv}.}
\end{figure}

A DES is characterized by a roughening exponent $\zeta$  that describes
how the displacement from a flat configuration grow with the distance $L$ between two points $u(L) \sim L^{\zeta}$. For a line in a two
dimensional space, three roughness exponents can be easily defined. The equilibrium roughness exponent corresponding to zero applied force is $\zeta_{eq} = 2/3$. If the line is moving very fast $F \gg F_c$ it averages over disorder and the system is thus equivalent to a pure thermal system. In one dimension, thermal fluctuations alone lead to a roughness exponent $\zeta_{th} = 1/2$. Finally, at zero temperature, if the line is right
at the critical force, it is characterized by the roughness exponent $\zeta_{dep} \simeq 1.25$ \cite{rosso_hartmann}. The analogy with a standard critical phenomenon would thus suggest that an observation of the roughness of the line (in the moving frame in which the line is at rest) would lead to a divergent lengthscale on both side of the transition separating the various roughness regimes as depicted on Fig.~\ref{fig:regimes}.
However this picture disagreed with a functional renormalization group analysis of the small force (creep) regime, that was suggesting that $\zeta_{eq}$ should be observed at short, not large lengthscales \cite{chauve_creep}.

Given the difficulty to analyze this question analytically or by using the standard numerical algorithms (Langevin dynamics for example) due to the
extremely long computational times needed close to the depinning, we developed a new algorithm to tackle such a problem. Details on the algorithm can be found in \cite{kolton_depinning_zerot,kolton_depinning_zerot_long}. The basis idea is to use the fact that for very small temperatures, the system will spend most of its time in the configuration that has the largest escape barrier, since the escape time becomes infinitely longer than for all the other configurations. The steady
state properties of the system are thus dominated by this particular configuration. This is the generalization for dynamics of the case of equilibrium thermodynamics where the partition function of the system is dominated at low temperatures by the ground state of the system. Our algorithm allows
by enumeration to find such a configuration. Given the fact that the algorithm does not simulate the (extremely slow) dynamics close to the depinning, it can obtain quite quickly this configuration without suffering from the slowdown of the dynamics at low temperatures.

\begin{figure}
\includegraphics[width=8cm]{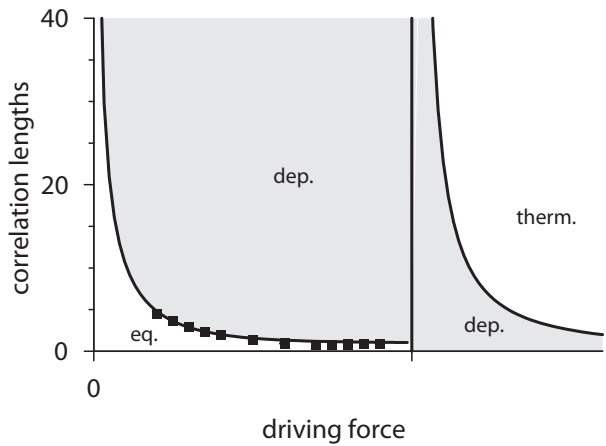}
\caption{\label{fig:nodiv} Correlation length (squares) below the depinning transition, as computed by the algorithm described in the text. The line is a guide to the eye. Contrarily
to what happens for $F>F_c$ there is \emph{no} divergent lengthscale showing up in the steady state properties of the line for $F < F_c$, in contrast with the naive picture
that could be intuited from the analogy with a standard critical phenomenon. After \cite{kolton_depinning_zerot}.}
\end{figure}

\begin{figure}
\includegraphics[width=8cm]{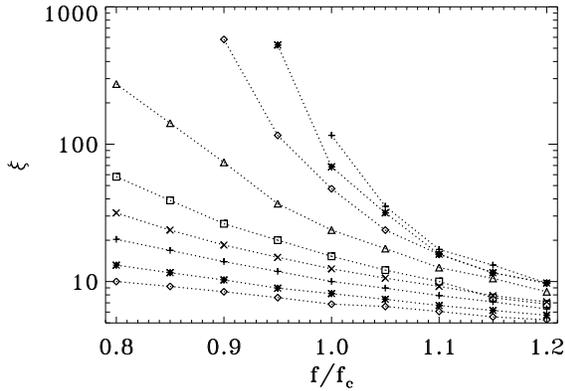}
\caption{\label{fig:molec} Characteristic length $\xi$ as a function
of the force $f$ for different temperatures $T$ increasing from the top to the bottom ($T
= 0.0025,0.005,0.01,0.025,0.05,0.075,0.1,0.15,0.2$) as computed by Langevin dynamics.  $\xi$
grows monotonically with decreasing $f$.  This is incompatible
with the possibility of having a divergent steady-state length below threshold and is in full agreement with the results obtained by a different algorithm
and shown in Fig.~\ref{fig:nodiv}.}
\end{figure}

The results are sketched in Fig.~\ref{fig:nodiv}.
They show clearly that \emph{no} divergent lengthscale exists in the steady state properties of the line for $F < F_c$, in direct contrast with
a naive extrapolation of our intuition on standard critical phenomena. In order to check that this result was no an artefact or our algorithm and that no subtle inversion of limits $T \to 0$ and $F \to F_c$ could affect this result, we also spend considerable time recovering the crossover length by more conventional Langevin dynamics simulations. The results are shown in Fig.~\ref{fig:molec}
and fully confirm the above results and the absence of divergent lengthscale on the lower side of the transition.

These results clearly shows that the intuition based on standard critical phenomena should be taken with a grain of salt and poses the problem
of the theoretical description of the depinning transition at finite temperatures. As for the case of the thermal rounding, although this
information is in principle encoded in the FRG equations at finite temperature, extracting the physical behavior from these equations remain
a very challenging task.

\section{Conclusion}

We have analyzed the effect of finite temperature on the depinning transition of a disordered elastic system. We have shown that at finite temperature
the depinning transition is thermally rounded. The thermal rounding exponent $\psi \simeq 0.15$ needs a theoretical explanation, since
for the moment the only formula which seems to fit quantitatively is an unjustified extrapolation of the static critical rounding exponents for a standard critical phenomenon. Although the existence of thermal rounding would be in good agreement of the extrapolation of our intuition
on critical phenomena to the depinning transition, we have also shown that the depinning transition is not accompanied by a divergent lengthscale
for $F < F_c$ that would have bearings on the steady state properties of the line. This is in stark contrast with what happens in a normal critical
phenomena, and directly prompts for a good description of the depinning transition.

\section{Acknowledgements}
This work was supported in part by the Swiss FNS, under MaNEP and division II. SB and AK acknowledge financial support from CNEA and CONICET.










\end{document}